\documentclass[prl,twocolumn,showpacs,amsmath,letter,unsortedaddress]{revtex4}
\usepackage{graphicx,verbatim}

\newcommand{\Journal}[4]{#1 \textbf{#2}, #3 (#4)}

\begin{document}
	
\title{Cooperative Multiscale Aging in a Ferromagnet/Antiferromagnet Bilayer}
\author{S. Urazhdin}\email[]{sergei.urazhdin@emory.edu}
\affiliation{Department of Physics, Emory University, Atlanta, Georgia 30322, USA}

\begin{abstract}
We utilize anisotropic magnetoresistance to study temporal evolution of the magnetization state in epitaxial Ni$_{80}$Fe$_{20}$/Fe$_{50}$Mn$_{50}$ ferromagnet/antiferromagnet bilayers. The resistance exhibits power-law evolution over a wide range of temperatures and magnetic fields, indicating that aging is characterized by a wide range of activation time scales. We show that aging is a cooperative process, i.e. the magnetic system is not a superposition of weakly interacting subsystems characterized by simple Arrhenius activation. The observed effects are reminiscent of avalanches in granular materials, providing a conceptual link to a broad class of critical phenomena in other complex condensed matter systems.
\end{abstract}


\maketitle

In bilayers of materials with different lattice parameters, structural frustration can result in dislocations or even amorphous interlayers~\cite{Chen2005}.  Similarly, magnetic frustration can be expected at interfaces between materials with different magnetic orders~\cite{Toulouse77}. The origin of the frustration is the random effective field experienced by both materials due to their exchange interaction across the interface that is generally not atomically smooth. In particular, some of the unusual magnetic properties exhibited by bilayers of antiferromagnets (AF) and ferromagnets (F) have been attributed to the magnetic domain walls that are formed to reduce the interfacial exchange energy~\cite{Mauri87,Malozemoff87}, or even disordered spin states near the F/AF interface~\cite{Schlenker86,Yamada2007}. After almost 60 years of extensive research, fundamental understanding of F/AF bilayer systems remains elusive. Besides theoretical challenges in describing the effective exchange fields at the F/AF interfaces, common experimental approaches, such as the hysteresis loop measurement, can lead to irreversible changes of the magnetic configuration, thus obscuring the essential signatures of frustrated systems such as aging~\cite{Fischer91,Vincent2007}. Therefore, nonperturbative real-time characterization methods may be needed to provide insight into the properties of these systems.

We utilized anisotropic magnetoresistance (AMR) to characterize the evolution of the magnetization state in F/AF bilayers without perturbing the system. Our measurements reveal power-law relaxation over a wide range of temperature $T$, indicating multiple scales of activation energies, and activation times spanning an estimated range of at least seven orders of magnitude. These results are not affected by the variations of field $H$, demonstrating that activation occurs in the AF layer. The observed power-law form of relaxation is also independent of the prior magnetic aging history, indicating that aging is a cooperative process; it cannot be described in terms of independent activation barriers. These results provide an unprecedented insight into the magnetism in F/AF bilayers, linking them to other complex and frustrated systems that exhibit cooperative aging phenomena.

Our samples were deposited by high-vacuum sputtering on (0001)-oriented sapphire substrates annealed in air at $1300^\circ$~C to achieve atomically flat surface. An epitaxial (111)-oriented Pt(5) buffer layer was deposited at $550^\circ$~\cite{Parkin92}, followed by the F/AF bilayer Ni$_{80}$Fe$_{20}$(10)/Fe$_{50}$Mn$_{50}$($d$) deposited at room temperature to avoid interdiffusion of the magnetic interfaces. All thicknesses are in nanometers (nm). The bilayer was capped with SiO$_2$(20) to prevent oxidation. We have fabricated and studied several samples with thickness $d$ of FeMn ranging from $1$ to $3.5$~nm. Their magnetic properties, such as the temperature $T_B$ characterizing the onset of hysteresis loop asymmetry - the exchange bias (EB)~\cite{Meiklejohn56,Nogues99}, were consistent with the previous studies~\cite{Parkin90,Jungblut94,Offi2002,Urazhdin2005}. We focus on a sample with $d=2$~nm, which exhibited $T_B=140$~K within the range of temperatures $T=5-300$~K accessible in our magnetoelectronic measurements. We will show below that some of the signatures of EB extend far above $T_B$.

%%%%%%%%%%%%%%%%%%%%%%%%%%%%%%%%%%%%%%%%%%%%%%%%%%%%%%%%%%%%%%%%%%%
\begin{figure}
	%\vspace{+5mm}
	\includegraphics[width=3.35in]{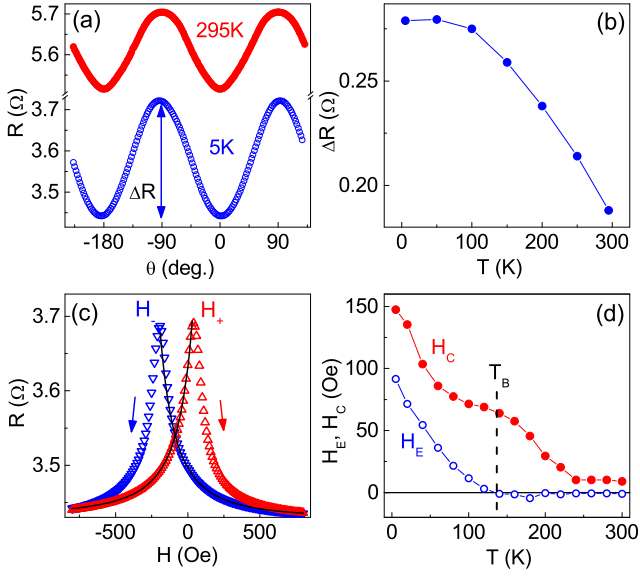}
	\caption{(Color online) Magnetoelectronic characterization of the F/AF bilayer: (a) Dependence of resistance on the in-plane orientation of the field, at $T=295$K, $H=100$~Oe (solid symbols), and $5$~K, $H=2$~kOe (open symbols), (b) Temperature dependence of magnetoresistance determined from $R(\theta)$ data obtained at $H=2$~kOe, (c) Symbols: magnetoelectronic hysteresis loop obtained at $5$~K, after 4 prior similar "training" loops, up(down) triangles are for increasing(decreasing) $H$. Curves: fits based on the effective exchange field model, as described in the text. Coercive fields $H_-$ and $H_+$ are labeled. (d) Temperature dependence of coercivity $H_C=(H_+-H_-)/2$ and exchange bias field $H_E=-(H_-+H_+)/2$.}\label{fig1}
	%\vspace{+5mm}
\end{figure}
%%%%%%%%%%%%%%%%%%%%%%%%%%%%%%%%%%%%%%%%%%%%%%%%%%%%%%%%%%%%%%%%%%%%%%%%%%%%

The magnetoelectronic characterization was performed in the four-probe van der Pauw geometry, using an ac current $I=0.1$~mA rms at frequency $f=1.3$~kHz, and lock-in detection of ac voltage $V$. The resistance $R=V/I$ exhibits a $180^\circ$-periodic sinusoidal dependence on the in-plane orientation of a sufficiently large in-plane field $H$ [Fig.~\ref{fig1}(a)], as expected due to the AMR of the Permalloy(Py)=Ni$_{80}$Fe$_{20}$ layer. The angle $\theta=0$ corresponds to the field orientation perpendicular to the current. The monotonic increase of the magnetoresistance $\Delta R=R(90^\circ)-R(0)$ with decreasing temperature $T$ is not affected by EB [Fig.~\ref{fig1}(b)], confirming that the magnetoresistance is determined entirely by the AMR of Py.

When $H$ is swept at $\theta=0$, the resistance exhibits sharp peaks at the coercive fields $H_+$, $H_-$ [Fig.~\ref{fig1}(c)]. To establish EB, we used the conventional procedure~\cite{Meiklejohn56,Nogues99} of cooling from room temperature $RT=295$~K in saturating field $H=500$~Oe, at $\theta=0$. The hysteresis loop becomes asymmetric below $T_B=140$~K, as illustrated in Fig.~\ref{fig1}(c) for $T=5$~K. We note that the magnetoresistance (MR) observed in the hysteresis loop is close to $\Delta R$ determined from the rotational AMR [Fig.~\ref{fig1}(b)], indicating that reversal occurs through the configuration of $\mathbf M$ almost homogeneously transverse to the field, consistent with a high spacial uniformity of the magnetic properties. The dependencies of both the EB field $H_E=-(H_1+H_2)/2$ and the coercivity $H_C=(H_2-H_1)/2$ on $T$ are consistent with other studies of EB systems based on FeMn~\cite{Parkin90,Jungblut94,Offi2002,Urazhdin2005}.

%%%%%%%%%%%%%%%%%%%%%%%%%%%%%%%%%%%%%%%%%%%%%%%%%%%%%%%%%%%%%%%%%%%
\begin{figure}
	%\vspace{+5mm}
	\includegraphics[width=3.35in]{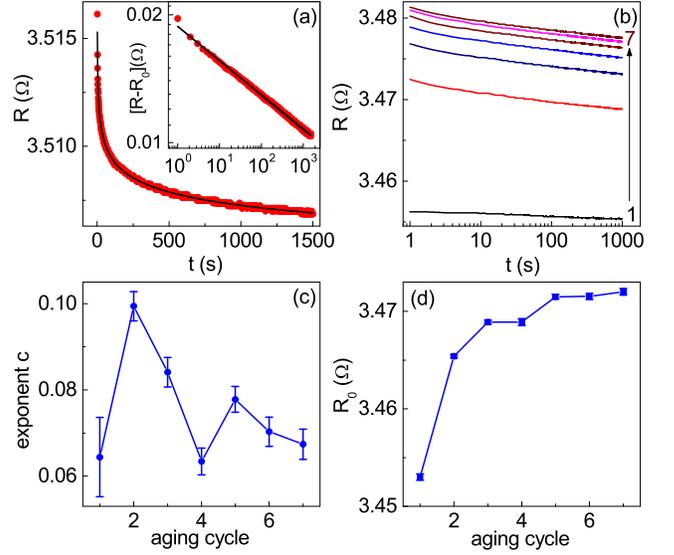}
	\caption{(Color online) Magnetic aging at $T=5$~K. (a) Symbols: time evolution of resistance $R$ at $H_f=-300$~Oe, after prior aging at $H_i=350$~Oe for $1000$~s. Curve: fitting with the power-law dependence $R(t)=R_0+At^{-c}$, with $R_0=3.496$~$\Omega$, $A=0.011$, $c=0.081$. Inset: $R-R_0$ vs $t$ plotted on the log-log scale. (b) Seven sequential aging cycles at $H_f=-500$~Oe, as labeled, each preceded by aging at $H_i=350$~Oe for $1$~Ks. (c,d) Dependence of the exponent $c$ (c) and the asymptotic resistance $R_0$ (d) on the aging cycle. The measurements were performed immediately after cooling at $H=500$~Oe from RT. The shown fitting error bars underestimate the uncertainty of the determined values, due to the additional errors from transients at small $t$.  
}\label{fig2}
	%\vspace{+5mm}
\end{figure}
%%%%%%%%%%%%%%%%%%%%%%%%%%%%%%%%%%%%%%%%%%%%%%%%%%%%%%%%%%%%%%%%%%%%%%%%%%%%

Our central result is the demonstration of cooperative multiscale aging over a wide range of $T$ and $H$. To observe aging, the field $H$ was ramped at a rate of $2$~kOe/s from the initial value $H_i$ above $H_+$ [or below $H_-$] to a final value $H_f$ below $H_-$ [or above $H_+$], and subsequently $R$ was recorded in $1$~s time increments, as illustrated in Fig.~\ref{fig2}(a) for $T=5$~K, $H_f=-300$~Oe.  The lock-in time constant was set to $100$~ms to minimize the effects of the instrumental bandwidth. We emphasize that both $H$ and $T$ were constant during aging, i.e. the recorded evolution was not perturbed by the measurement. Thermal activation in F/AF heterostructures has been extensively discussed in the context of granular systems, where the particle size is expected to set the energy scale $E_0$ for the Arrhenius-type exponential decay $R(t)=R_0+R_1exp[-t/\tau]$ over the characteristic time $\tau\propto exp[E_0/kT]$~\cite{Fulcomer72,Stiles99}. Here, $k$ is the Boltzmann constant. In our measurements, the evolution of $R(t)$ for $t$ up to about $50$~s could be well fitted with simple exponential decay. However, such fitting became inadequate at longer time scales. To remedy this discrepancy, one can assume a certain distribution of activation barriers, providing additional fitting parameters~\cite{Heijden98}. In the extreme limit, there is no characteristic activation energy scale, and consequently $R(t)$ does not exhibit a characteristic decay time. A similar situation is encountered at critical points in phase transitions, resulting in power-law dependencies of physical properties~\cite{Stanley71}. Indeed, power-law dependence $R=R_0+At^{-c}$ provided an excellent fit for all of our experimental $R(t)$ data, as illustrated by the curve in Fig.~\ref{fig2}(a) for data spanning three orders of magnitude in $t$. The power-law dependence was observed at temperatures from $5$~K to over $200$~K, indicating that the activation energies are spread over more than two orders of magnitude.

The aging curves depended on the previous aging history, reminiscent of the training effect - variations of the magnetic properties observed in sequential magnetic hysteresis loops of F/AF bilayers~\cite{Schlenker86,Hauet2006}. Figure~\ref{fig2}(b) shows seven aging curves acquired at $H_f=-500$~Oe. Each curve was measured after pre-aging at $H_i=300$~Oe over time $\Delta t_i=1$~Ks. The overall form of the dependence remains similar for different aging cycles, as confirmed by fitting with the power-law dependence, with exponent $c$ that exhibits only a modest irregular dependence on the aging cycle [Fig.~\ref{fig2}(c)]. The main difference between the consecutive aging curves is the overall increase of $R$. Since the value of $R$ itself evolves in time, the increase of $R$ in consecutive cycles can be characterized by its asymptotic value $R_0$, which exhibits a monotonic increase with the aging cycle number [Fig.~\ref{fig2}(c)]. This result demonstrates that the state asymptotically reached by the magnetic system after a single aging cycle is dependent on the magnetic history, similar to the kinetic trapping in glasses~\cite{Wales2003, Binder2005}. The data of Fig.~\ref{fig2} confirm the dependence of the magnetic properties, including aging characteristics, on the specific protocol used to prepare the magnetic system, which is well known both in the EB community and in studies of glassy systems. The results described below were obtained after multiple aging cycles, in the regime where the main aging characteristics stabilize [see Figs.~\ref{fig2}(c,d)]. We anticipate that studies of the dependence on the preparation protocol may provide insight into the aging mechanisms.

%%%%%%%%%%%%%%%%%%%%%%%%%%%%%%%%%%%%%%%%%%%%%%%%%%%%%%%%%%%%%%%%%%%
\begin{figure}
	%\vspace{+5mm}
	\includegraphics[width=3.35in]{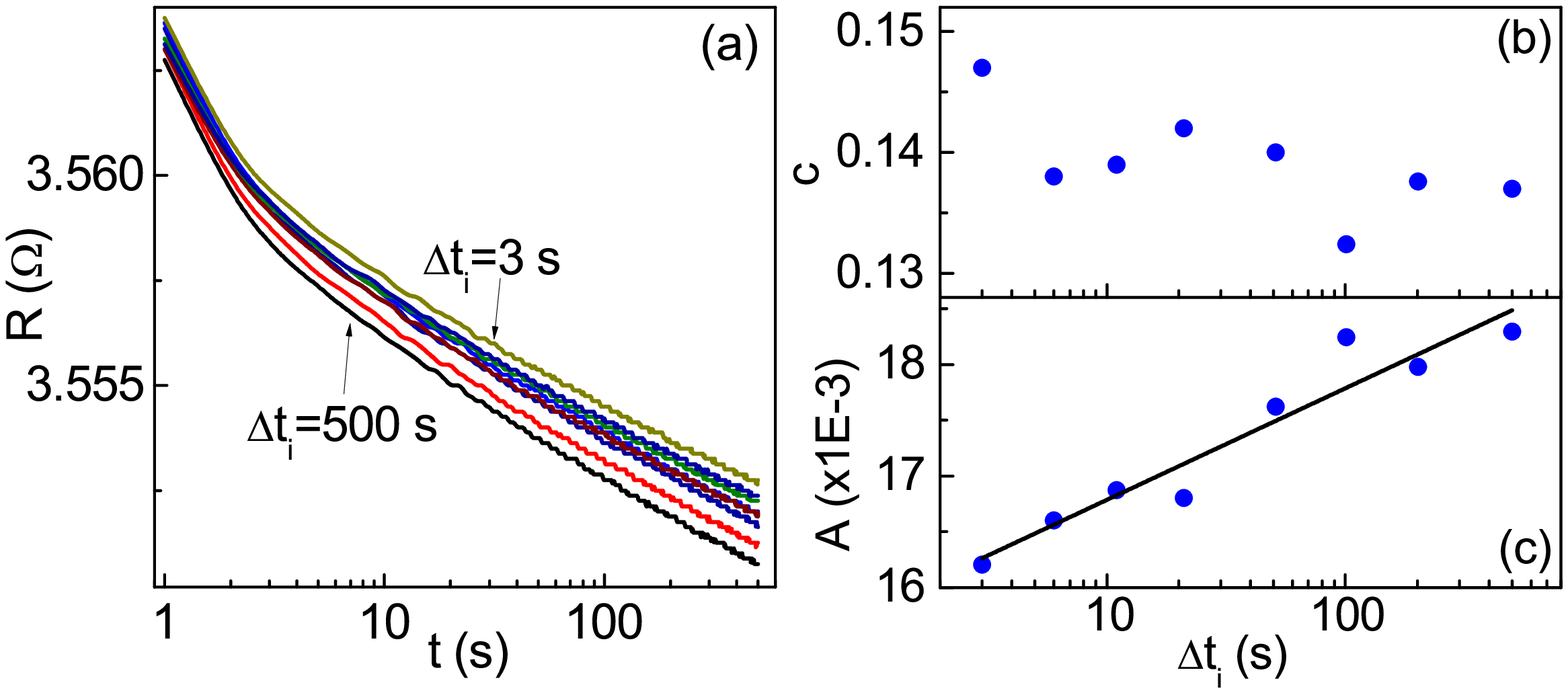}
	\caption{(Color online) (a) $R$ vs $t$ at $T=5$~K, $H_f=150$~Oe, after prior aging at $H_i=-300$~Oe over time interval $\Delta t_i=500$~s (bottom curve), $200$~s, $100$~s, $51$~s, $21$~s, $11$~s, $6$~s, and $3$~s (top curve). (b,c) Power-law exponent $c$ (b) and relaxation scale $A$ (c) vs $\Delta t_i$ (symbols), and a logarithmic fit to the data (line). Note the logarithmic scale for $\Delta t_i$.
	}\label{fig3}
	%\vspace{+5mm}
\end{figure}
%%%%%%%%%%%%%%%%%%%%%%%%%%%%%%%%%%%%%%%%%%%%%%%%%%%%%%%%%%%%%%%%%%%%%%%%%%%%

The dependence of $R_0$ on the aging history indicates that the system cannot be described in terms of independent activation barriers. To further test this conclusion, we performed aging measurements in which we varied only the time $\Delta t_i$ of pre-aging in the reversed state, from $500$~s down to $3$~s [Fig.~\ref{fig3}(a)]. The shape of the aging curves did not depend on $\Delta t_i$, and the decay exponent $c$ obtained from the power-law fitting remained the same within the approximately $10\%$ data spread [Fig.~\ref{fig3}(b)]. If the relaxation could be described by independent activation processes characterized by the barriers $E_n$ and the corresponding relaxation times $\tau_n$, $1\le n\le N$, then the subsystems with $\tau_n<\Delta t_i$ would become activated, while the subsystems with $\tau_n>\Delta t_i$ would not be activated. As a consequence, for small $\Delta t_i$ the aging curves would exhibit significantly smaller amplitude of decay at  $t>\Delta t_i$. Since the form of the experimental aging curves is independent of $\Delta t_i$, we conclude that aging involves cooperative processes coupling multiple energy scales, and cannot be described by the Arrhenius-type activation of individual weakly coupled subsystems.
	
While the exponent $c$ characterizing the aging curves remained independent of $\Delta t_i$, the decay scale $A$ exhibited a small but well-defined decrease by about $15\%$ when $\Delta t_i$ was decreased by two orders of magnitude [Fig.~\ref{fig3}(c)]. The values of $\Delta t_i$ below a few seconds are not experimentally accessible in our magnetoelectronic measurements. Nevertheless, one can generally expect that $A$ should vanish when $\Delta t_i$ becomes smaller than the shortest activation timescale in the system. The approximately logarithmic dependence in Fig.~\ref{fig3}(c) extrapolates to an intercept $A=0$ at $10^{-4}$~s. Thus, activation timescales likely span at least seven orders of magnitude, from $10^{-4}$~s to at least our characteristic measurement time of $10^3$~s. We note that the extrapolation of aging characteristics to $t=10^{-4}$~s does not lead to unphysical results. For instance, the total estimated variation of resistance, $R(t=10^{-4}s)-R_0=10^{4c}A$, did not exceed $0.12$~$\Omega$ in all of the aging experiments at $5$~K. This value is smaller than the full MR $\Delta R=0.28$~$\Omega$ at $5$~K.

%%%%%%%%%%%%%%%%%%%%%%%%%%%%%%%%%%%%%%%%%%%%%%%%%%%%%%%%%%%%%%%%%%%
\begin{figure}
	%\vspace{+5mm}
	\includegraphics[width=3.35in]{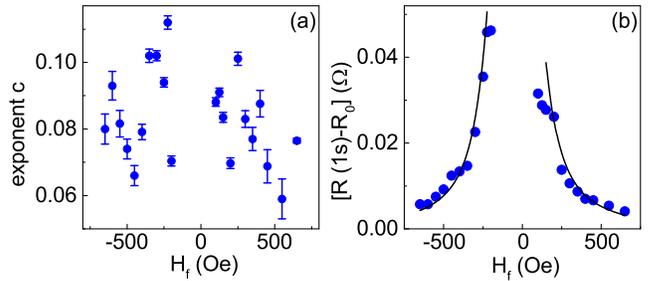}
	\caption{(Color online) Dependence of the aging characteristics on field, at $T=5$~K. (a) The power-law exponent $c$ determined from fitting the $R$ vs $t$ data. Error bars show the fitting uncertainty. (b) Symbols: experimental dependence of the total relaxation amplitude $R(t=1s)-R_0$ on field. Lines: fitting based on the effective exchange field model, as discussed in the text, using the data for $|H|>200$~Oe.
	}\label{fig4}
	%\vspace{+5mm}
\end{figure}
%%%%%%%%%%%%%%%%%%%%%%%%%%%%%%%%%%%%%%%%%%%%%%%%%%%%%%%%%%%%%%%%%%%%%%%%%%%%

Both F and AF can contribute to aging in F/AF bilayers. The magnetic anisotropy and/or defects in F and/or AF can define the energy barriers for the magnetic evolution, which can proceed either by the uniform reversal in small magnetic grains, or by F or AF domain wall motion in continuous films. To establish the relative contributions of the two magnetic layers, we determined the dependence of aging characteristics on the field $H_f$. The Zeeman energy contribution should result in the exponential dependence of activation in the F layer on $H_f$, while the corresponding dependence for AF should be weak. Figure~\ref{fig4}(a) summarizes the values of the exponent $c$ determined from aging at different $H_f$ ranging from $-650$~Oe to $650$~Oe. The magnetoelectronic signatures of aging became too small for reliable measurements at $|H_f>650|$~Oe [see Fig.~\ref{fig4}(b)]. The values of $c$ exhibit random variations around the average $c=0.084$, and no correlation with $H_f$. Based on this result, we conclude that aging involves activation processes in AF that affect the magnetization $M$ of Py only through exchange coupling at the interface and are not directly influenced by $H$. The error bars in Fig.~\ref{fig4}(a) reflect only the fitting uncertainty, and do not account for the additional errors caused by the transient effects at short times scales, which are caused by the limited bandwidth of both the electromagnet power supply and the lockin amplifier. An additional measurement time error is caused by the onset of aging while the field is still being ramped. 

While the form of the aging curves was independent of $H_f$, the overall scale of the resistance decay rapidly decreased with increasing magnitude of $H_f$, as shown by symbols in Fig~\ref{fig4}(b). Here, we plot the difference between the first measured resistance value at $R(t=1s)$ and its asymptotic value $R_0$, which, as we shall see below, is more convenient than the scale $A$ for the quantitative analysis of relaxation. The dependence in Fig.~\ref{fig4}(b) is consistent with our conclusion that aging occurs in the AF layer. We can describe the exchange interaction of F with AF by an effective exchange field $\mathbf H'$ with average components $H'_\parallel$, $H'_\perp$ in the direction of $H$ and perpendicular to it, respectively. Both of these components vary over time due to the AF aging, resulting in variations of the angle $\phi$ formed by the Py magnetization $M$ relative to $H$, according to $\phi\approx H'_\perp/(H+H'_\parallel)$, and the corresponding variations of resistance 
\begin{equation}\label{eq:fit}
R=R_{min}+\frac{\Delta R}{2}\left[\frac{H'_\perp}{H+H'_\parallel}\right]^2
\end{equation}
where $R_{min}$ is the resistance minimum at $\phi=0$. Our model is supported by the excellent agreement of the fit based on Eq.~(\ref{eq:fit}) with the quasi-static measurements of $R$ vs $H$ [solid curves in Fig.~\ref{fig1}(c)], yielding $H'_\perp=148.5\pm.5$~Oe from two independent fits of both hysteresis loop branches up to the switching points. Fitting the same curves beyond the switching point is less meaningful, because of the aging that occurs concurrently with the field sweep. Using the form of Eq.~(\ref{eq:fit}) to analyze the dependence of the relaxation magnitude on $H_f$ in Fig.~\ref{fig4}(b), we obtain a good fit for all the $|H|>200$~Oe data, with a single set of fitting parameters $H'_\parallel=50$~Oe, $\Delta H'_\perp=105$~Oe [curves in Fig.~\ref{fig4}(b)]. Here, $\Delta H'_\perp$ is the overall reduction of the effective transverse field due to aging between $t=1$~s and $\infty$. Thus, the dependence of the aging curves on $H_f$ can be explained entirely by the effect of $H$ on $M$, while its direct effect on aging is negligible.

%%%%%%%%%%%%%%%%%%%%%%%%%%%%%%%%%%%%%%%%%%%%%%%%%%%%%%%%%%%%%%%%%%%
\begin{figure}
	%\vspace{+5mm}
	\includegraphics[width=3.35in]{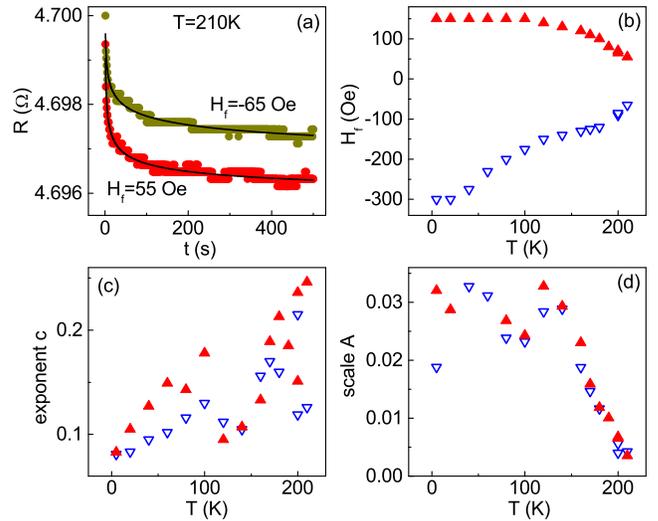}
	\caption{(Color online) Dependence of aging characteristics on temperature. (a) Symbols: measured $R$ vs $t$ at $T=210$~K, $H_f=-65$~Oe and $H_f=55$~Oe, as labeled. Curves: fits with the power-law dependence, and power-law exponents $c=0.126$ and $0.246$. (b) $H_f$ used in the measurements of aging at different $T$. (c,d) Power-law exponent $c$ (c) and scale $A$ (d) vs $T$, obtained from fitting of the aging data such shown as in panel (b). Filled (open) symbols are for $H_f>0$ ($H_f<0$).  
	}\label{fig5}
	%\vspace{+5mm}
\end{figure}
%%%%%%%%%%%%%%%%%%%%%%%%%%%%%%%%%%%%%%%%%%%%%%%%%%%%%%%%%%%%%%%%%%%%%%%%%%%%

Behaviors similar to those discussed above for $T=5$~K were observed at higher $T$, extending significantly above the blocking temperature $T_B=140$~K [see Fig.~\ref{fig5}(a) for aging curves and their power-law fits at $210$~K]. To quantitatively characterize the dependence on $T$, aging was measured at the values of $H_f$ adjusted so that the corresponding resistance in the hysteresis loop was approximately at $20\%$ of $\Delta R$ above the minimum [Fig.~\ref{fig5}(b)]. The power-law exponent $c$ exhibited an overall increase from $0.08$ at $5$~K to about $0.2$ around $200$~K [Fig.~\ref{fig5}(b)]. This trend is superimposed with increased random variations of $c$, which can be correlated with the decrease of the relaxation scale $A$ above $T_B$ [Fig.~\ref{fig5}(d)], preventing reliable measurements of aging at $T>210$~K. The increase of $c$ with increasing $T$ is consistent with the larger relative contribution of fast activation processes, and a smaller contribution of slow processes. Nevertheless, the power-law form of aging curves is retained up to $210$~K [Fig.~\ref{fig5}(a)], indicating non-negligible contribution from the activation times $\tau\approx 1000$~s. The decrease of $A$ at high $T$ can be attributed to the decreasing volume of AF involved in aging on the experimentally accessible timescales.

Summarizing our main findings, magnetoelectronic measurements reveal aging in epitaxial thin-film F/AF=Py/FeMn bilayers over a wide range of temperatures. The aging curves exhibit simple scaling with the applied magnetic field, demonstrating that the activation processes are confined to the AF layer.  Aging is characterized by the power-law dependence on time and a weak dependence on temperature, indicating a wide range of the activation energies, and a range of the activation times that at $T=5$~K is estimated to extend over at least seven orders of magnitude, from $10^{-4}$~s to at least the longest measurement time of $10^3$~s. Finally, the form of the aging curves is independent of the magnetic pre-aging history, indicating that activation processes are cooperative, in other words the magnetic system cannot be described as a superposition of weakly-coupled subsystems activated according to the Arrhenius law. 

The observed power-law evolution is reminiscent of the self-organized criticality and the associated avalanche dynamics~\cite{Bak87}. In ferromagnets, such avalanches of magnetic domain reversal are observed as the Barkhausen noise~\cite{Urbach95}. The emerging physical picture for aging in AF is that of stable AF regions interfaced with regions that can be either stable or activated, depending on the stability of their AF environment. Activation of a certain region can activate or deactivate other neighboring AF regions due to their exchange interaction, which can result in avalanches of AF activation. According to the picture of self-organized criticality~\cite{Bak87}, at long times the AF is expected to asymptotically form "minimially stable clusters", which can be  irreversibly perturbed by small variations of temperature or changes of the magnetic configuration of F. This picture implies a correlation between the effects of the magnetic and the thermal history. Testing this correlation can elucidate the mechanisms of cooperativity.

The geometry of the activated AF clusters (and by extension the stable AF regions) is likely fractal, since aging phenomena are driven by the random effective exchange field. Our hypothesis is supported by the apparent lack of the characteristic activation energy scale, since the latter is determined by the AF anisotropy and scales with the volume of the activated regions. Further spatially resolved studies, e.g. based on x-ray dichroism microscopy of AF and/or F layers will likely elucidate the spatial characteristics of activation [see  Ref.~\cite{Benassi2014} for microscopic images quite conducive of this physical picture]. Our results may have significant implications for other F/AF heterostructures based on single-crystal and even polycrystalline AF materials, where crystallinity may provide a natural limit for the geometry of the activated AF clusters and cooperativity, but some signatures of the behaviors described above are likely retained.  F/AF bilayers may also represent a controllable (by means of the magnetic field and temperature) model system that can provide insight into other condensed matter systems exhibiting complex critical phenomena. For instance, the random effective exchange field experienced by the AF can provide a straightforward implementation for the classic Imry-Ma problem of random-field magnetism~\cite{Imry75,Proctor2014}.

We thank Eric Weeks for helpful discussions. This work was supported by the NSF grant DMR-1504449.

\end{document}